\documentclass[reprint,amsmath,amssymb,aps,pra,longbibliography]{revtex4-2}

\usepackage{graphicx}
\usepackage{epsfig}
\usepackage{epstopdf}
\usepackage{subfigure,color}
\usepackage{dcolumn}
\usepackage{bm}
\usepackage{lineno}
\usepackage{amsmath, amssymb}
\usepackage{accents}
\usepackage{lipsum}
\usepackage{multirow}
\usepackage[usestackEOL]{stackengine}
\usepackage{tabularx}
\usepackage{makecell}
\usepackage{xcolor}
\usepackage{dcolumn}% Align table columns on decimal point
\usepackage{bm}% bold math
\usepackage{booktabs}
\def\e{\begin{equation}}
\def\f{\end{equation}}
\def\_#1{{\bf #1}}
\def\/#1{_{\rm #1}}
\def\.{\cdot}

\begin{document}
	
	\preprint{APS/123-QED}
	
	%	\title{Wave isolation in Space-Time Modulated Surface Impedance}
	
	%\title{Multifunctional nonreciprocal meta}
	
	\title{ Theory and Design of Multifunctional Space-Time Metasurfaces}
	
	\author{Xuchen Wang$^1$, Ana~D\'{i}az-Rubio$^1$,  Huanan Li$^2$,  Sergei A. Tretyakov$^1$, and Andrea Al\`{u}$^2$}
	
	\affiliation{$^1$Department of Electronics and Nanoengineering, Aalto University, P.O.~Box 15500, FI-00076 Aalto, Finland}
	\affiliation{$^2$Photonics Initiative, Advanced Science Research Center, City University of New York, New York, USA}
	
	\date{\today}

	\begin{abstract}
		Integrating multiple functionalities into a single metasurface is becoming of great interest for future intelligent communication systems.  
		While such devices have been extensively explored for reciprocal functionalities, in this work, we integrate a wide variety of nonreciprocal applications into a single platform.
		The proposed structure is based on spatiotemporally modulated impedance sheets supported by a grounded dielectric substrate. 
		We show that, by engineering the excitation of evanescent modes, nonreciprocal interactions with impinging waves can be configured at will.
		We demonstrate a plethora of nonreciprocal components such as wave isolators, phase shifters, and circulators, on the same metasurface. 
		This platform allows switching between different functionalities only by modifying the pumping signals (harmonic or non-harmonic), without changing the main body of the metasurface structure. 
		This solution opens the door for future real-time reconfigurable and environment-adaptive nonreciprocal wave controllers.
	\end{abstract}
	
	\maketitle
	
	\section{Introduction}
	
	The next generations of communication systems will need multifunctional devices that adapt to different usage and environment  requirements, enhancing user experiences and providing better information exchange.
	In this emerging paradigm, due to their integration capabilities and the unprecedented opportunities for controlling waves, metasurfaces are positioned as the platform of choice to implement multifunctional responses.
	By varying the local properties of each composing element in space and time, one can create metasurfaces with different scattering responses. Metasurfaces can be treated as multi-port devices that are able to simultaneously control waves incoming and outgoing from/to a number of propagation directions, for example, waves coming from oppositely tilted angles \cite{asadchy2017flat,wang2018extreme}. 
	Following the classical notation of multi-port networks, the properties of a metasurface can be modeled by the scattering matrix, where each element $S_{ij}$ represents the ratio of the flux amplitudes between the outgoing waves to $i$-port and the incoming wave from $j$-port.
Assume that there is no frequency transformations during scattering, the functionalities of metasurfaces  can be classified into two groups: reciprocal, characterized by symmetric scattering matrices $(\overline{\overline S}^T=\overline{\overline S})$, and non-reciprocal, where scattering matrices are asymmetric $(\overline{\overline S}^T \neq \overline{\overline S})$ 
	
	Reciprocal metasurfaces have been intensively studied in the last decade.
	In practice, one can make use of {\it gradient} or {\it space-modulated} metasurfaces for tailoring the scattering properties between different ports and design anomalous reflectors \cite{diaz2017generalized,asadchy2016perfect,ra2017metagratings}, beam splitters, multichannel retroreflectors \cite{asadchy2017flat} or asymmetric absorbers \cite{wang2018extreme}.
	Recent advances in reciprocal metasurfaces have shown that one metasurface can perform a multitude of functionalities, switching from one application to another by reconfiguring electromagnetic parameters of the individual meta-atoms. For example, in \cite{liu2019intelligent}, independent tunability of resistance and reactance of each unit cell using embedded tunable circuits has been discussed. In this way, by locally modifying the impedance of each unit cell, the metasurface can act as a perfect absorber or anomalous reflector without changing the main body of the metasurface structure.
	
	Reciprocity imposes certain restrictions in the design of metasurfaces,  and there are practical applications where it is necessary to go beyond them.
	Traditionally, this goal has been achieved with magneto-optical materials, such as ferrites \cite{adam2002ferrite} and graphene \cite{tamagnone2016near,crassee2011giant,sounas2012gyrotropy}, biased by external static or slowly-varying magnetic fields.  
	But the use of bulky magnets makes these devices incompatible with integrated technologies. In addition, the weak magnetic response and high losses in terahertz and optical ranges further inhibit their use in future optoelectronic and nanophotonic systems. 
	One magnetless alternative is to embed nonlinear materials in resonant cavities \cite{sounas2018broadband,mahmoud2015all,d2019nonlinearity}. By creating spatial asymmetries in the cavity, the distribution of electromagnetic fields can be made significantly different for opposite illuminations.  However, this solution is restricted to strong excitation intensities, poor isolation, and nonsimultaneous excitation from different ports \cite{shi2015limitations,sounas2018fundamental}.
	
	In recent years, dynamic modulation of material properties has attracted considerable attentions in both physics and engineering communities, due to its great potentials to induce extraordinary wave phenomena~\cite{Ptitcyn2019time,mirmoosa2019time,sounas2017non,yuan2015three,buddhiraju2020photonic,koutserimpas2018nonreciprocal,liu2018huygens}. Time-varying materials break Lorentz reciprocity in linear and magnet-free devices. Spatiotemporal modulation  can impart synthetic linear or angular momentum in the operating systems that emulate high-speed mechanical motion, and therefore produce Doppler-like nonreciprocity \cite{mock2019magnet,hadad2016breaking,taravati2020space}.  
% 	Early works on space-time periodic media appeared in the middle of last century  \cite{cassedy1965waves,oliner1961wave,cassedy1963dispersion,cassedy1967dispersion}. While those efforts were mainly concentrated on the computation of scattering harmonics served for the study of parametric amplification, little attention was paid to nonreciprocal effects in such media and to their potential use. 
% 	During the last decade, the idea of space-time modulation of material properties has been revitalized and brought to applications for the realization of nonreciprocal devices. 
% 	In  \cite{yu2009complete,lira2012electrically,yu2009integrated},  optical isolation in photonic waveguides has been  demonstrated in both theory \cite{yu2009complete} and experiment \cite{lira2012electrically}, with a traveling-wave refractive index modulation.
% 	In \cite{hadad2015space}, one-way transmission of propagating waves has been achieved with the modulation of surface capacitance in space and time.
	Based on this principle, many interesting phenomena have been explored 	\cite{shaltout2019spatiotemporal},
% 	Afterwards, more possibilities and interesting phenomena based on different modulation schemes have been explored
such as wave isolation \cite{yu2009complete,lira2012electrically,yu2009integrated,correas2018magnetic,qin2016nonreciprocal,chamanara2017optical,taravati2017nonreciprocal,taravati2017self,hadad2015space} and circulation \cite{shi2017optical,estep2016magnetless}, one-way beam splitting \cite{taravati2019dynamic}, frequency conversion \cite{salary2018time,salary2018electrically,hadad2019space}, and nonreciprocal antennas \cite{hadad2016breaking,taravati2017mixer, correas2016nonreciprocal,zang2019nonreciprocal}. However, the nonreciprocal applications are usually fixed and unadaptable to changing environments or application requirements.
While the existing works mostly focus on a specific nonreciprocal effect realized with a fixed modulation scheme, in this paper, we show that it is possible to develop robust and general means for reconfigurable nonreciprocal functionalities on a metasurface platform. 
	
	In particular, we demonstrate that by altering the space-time modulation functions it is possible to implement various canonical nonreciprocal devices on a universal platform. 
	This multifunctional platform is based on a tunable surface impedance sheet supported by a metallized substrate. We assume that the resistance and reactance of the surface impedance can be independently and locally controlled in real time \cite{liu2019intelligent}.
	We first develop a generalized circuit theory to characterize the scattering harmonics of surface impedances modulated with arbitrary traveling waveforms (Section~\ref{Sec: Theory}). The theoretical formulation allows not only direct computations of scattering harmonics for a known modulated surface impedance, but also finding the optimal surface impedance that ensures excitation of a prescribed set of scattering harmonics.
	By incorporating evanescent fields into the picture, the optimized surfaces exhibit strong and controllable nonreciprocity. 
	Based on this principle, we assign optimal modulation functions to a single metasurface with fixed dimensions and demonstrate possibilities to realize isolators, phase shifters, quasi-isolators, and circulators (Section~\ref{Sec: Functionalities}). 
	In addition, we propose viable
	practical implementation schemes at different frequency ranges, showing realizability of the proposed multifunctional and adaptable platform (Section~\ref{Sec: implementation}).

	\section{Generalized Circuit Theory for Space-Time Modulated Surfaces} \label{Sec: Theory}

\begin{figure}[h]
		\centering
		\subfigure[]{\includegraphics[width=0.9\linewidth]{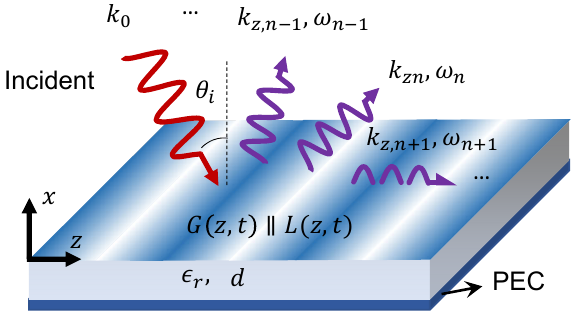}
			\label{fig:structure}} 
		\subfigure[]{\includegraphics[width=0.9\linewidth]{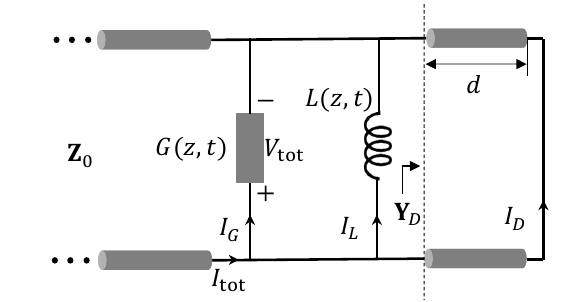}
			\label{fig: circuit model}}
		\caption{(a) General scattering scenario of space-time modulated impenetrable surface and (b) its equivalent circuit model. }\label{fig:structure_circuit}
\end{figure}

In this section, based on the surface impedance model, we present the theoretical formulation that describes a reconfigurable platform for nonreciprocal devices. We consider the structure shown in Fig.~\ref{fig:structure}: an electrically thin sheet mounted on a grounded substrate.
The main goal is to allow arbitrary reconfiguration of the nonreciprocal functionality using spatiotemporal modulations of the electric response of the sheet.

For a general impedance sheet, its electrical properties can be defined by the parallel connection of a conductance, $G(z,t)$,  and an inductance, $L(z,t)$, which are periodically modulated in space and time [see Fig.~\ref{fig:structure}] with spatial frequency $\beta_{\rm M}=2\pi/D$ ($D$ is the spatial period of modulation functions) and temporal frequency $\omega_{\rm M}=\beta_{\rm M}=2\pi/T$ ($T$ is the temporal period of modulation functions). 
We assume that $G(z,t)$ and $B(z,t)=1/L(z,t)$ are modulated with an arbitrary traveling waveform (the inverse of inductance, $B(z,t)$, is introduced in this paper for mathematical simplicity), and therefore they can be expressed as a sum of Fourier series \cite{suppl,elnaggar2019generalized},
\begin{equation}
\Psi(z,t)=\sum_{m=-\infty}^{+\infty}\psi_{m} e^{-jm(\beta_{\rm M} z-\omega_{\rm M}t)},
\label{Eq: travelling_wave_conductance}
\end{equation}
 where  $\Psi$ denotes either $G$ or  $B$, and $\psi_{m}$ represents  $g_m$ and $b_m$ which are the harmonic coefficients of $G(z,t)$ and $B(z,t)$, respectively. 
 It is important to notice that $\psi_m^*=\psi_{-m}$ to ensure that both $G(z,t)$ and $L(z,t)$ are real-valued  functions.

% To obtain the general expressions of the travelling-wave modulation, let us first assume that both $G$ and $B$ are spatially varying with periodicity $D$. Generally, these periodic functions can be expanded in Fourier series
% \begin{equation}
%   % \Psi(z)=\sum_{m=-\infty}^{+\infty}\psi_{m} e^{-jm\beta_{\rm M} z}, \quad \Psi\in[G,B], \psi\in[g,b]
%   \Psi(z)=\sum_{m=-\infty}^{+\infty}\psi_{m} e^{-jm\beta_{\rm M} z}, 
% 	\label{Eq: one dimensional_susceptance}
% \end{equation}
% where  $\Psi(z)$ denotes either $G(z)$ or  $B(z)$, $\psi_{m}$ represents  $g_m$ and $b_m$ which are the harmonic coefficients of $G(z)$ and $B(z)$, respectively,  and $\beta_{\rm M}=2\pi/D$ is the modulation wavenumber. It is important to notice that $\psi_m^*=\psi_{-m}$ to ensure that both $G(z)$ and $L(z)$ are real-valued  functions. This expansion allows engineering the conductance and the inductance with arbitrary periodic waveforms.

% Next, we assume that the waveforms $G(z)$ and  $B(z)$ are moving along the $z$-axis with a constant velocity $v_{\rm M}=D/T$ ($T$ denotes the temporal periodicity). In this case, $G$ and $B$ are periodical functions of both space and time, and their Fourier expansions can be obtained by replacing $z$ in Eq.~ (\ref{Eq: one dimensional_susceptance}) by $z-v_{\rm M}t$:
% \begin{equation}
% \Psi(z,t)=\Psi(z-v_{\rm M}t)=\sum_{m=-\infty}^{+\infty}\psi_{m} e^{-jm(\beta_{\rm M} z-\omega_{\rm M}t)},
% \label{Eq: travelling_wave_conductance}
% \end{equation}
% where $\omega_{\rm M}=v_{\rm M}\beta_{\rm M}=2\pi/T$ is the modulation frequency. 

 Then, if a plane wave (wavenumber $k_0$ and frequency $\omega_0$) impinges on the structure at the incidence angle $\theta=\theta_{\rm i}$, according to the Floquet-Bloch theorem, 
%the stationary scattered fields have the form,
% \begin{equation}
% K(z,t)=e^{-j(k_zz-\omega_0t)}P(z^\prime) \label{eq: floquet_space_time}
% \end{equation} 
% with $k_z=k_0\sin\theta_{\rm i}$ being the tangential wavenumber and $z^\prime=z-v_{\rm M}t$ \cite{elnaggar2019generalized}. Here, $K$ represents both electric, $E$, and magnetic, $H$, fields.  
% Notice that $P(z^\prime)$ is a periodical function with the spatial periodicity $D$, $P(z^\prime)=P(z^\prime+nD)$, and can be expanded in Fourier series: 
% \begin{equation}
% P(z^\prime)=P(z-v_{\rm M}t)=\sum_{n=-\infty}^{+\infty}p_{n} e^{-jn\beta_{\rm M}(z-v_{\rm M}t)}. \label{Eq:P(z)}
% \end{equation}
% After substituting Eq.~(\ref{Eq:P(z)}) into Eq.~(\ref{eq: floquet_space_time}) the scattered fields can be written as
% \begin{equation}
% K(z,t)=\sum_{n=-\infty}^{+\infty}p_{n}e^{-j[(k_z+n\beta_{\rm M})z-(\omega_0+n\omega_{\rm M})t]}.\label{eq: scattered fields}
% \end{equation} 
%From Eq.~(\ref{eq: scattered fields}), we can see that
the scattered fields contain an infinite number of Floquet harmonics. 
The $n$-th harmonic propagates with the transverse wavenumber $k_{zn}=k_z+n\beta_{\rm M}$ and has the frequency $\omega_n=\omega_0+n\omega_{\rm M}$ \cite{suppl}.
% Knowing the general expressions for the scattered harmonics, we use the mode-matching method to calculate their amplitudes and phases \cite{hwang2012periodic}.
We write the tangential components of the total electrical and magnetic fields on the surface as a sum of the incident wave and all the scattered harmonics:
	\begin{equation}
	\begin{split}
	E_{\rm tot}^t&=E_{\rm i}^t+E_{\rm s}^t=\sum_{n=-\infty}^{+\infty}E_n^t e^{-j(k_{zn} z-\omega_n t)}\\
	H_{\rm tot}^t&=H_{\rm i}^t+H_{\rm s}^t=\sum_{n=-\infty}^{+\infty}H_n^t e^{-j(k_{zn} z-\omega_n t)}. 
	\end{split} \label{Eq:field_expression}
	\end{equation}
The coefficients $E_n^t$ and $H_n^t$  can be uniquely determined by enforcing the boundary conditions on all the interfaces of the structure [see Fig.~\ref{fig:structure}].  However, this could be a cumbersome task, especially for structures with multiple constitutive layers.
To simplify the mathematical derivations, we analyze the metasurface using the transmission-line model shown in Fig.~\ref{fig: circuit model}.
In this model, the spatiotemporally varying sheet is represented by a shunt impedance formed by a parallel connection of an  inductor and a resistor. The metal-backed substrate is modeled as a section  of transmission line with the length equal to the substrate thickness. The infinitely extended transmission line on the left side of Fig.~\ref{fig: circuit model} represents free space, and its characteristic impedance, denoted as $\mathbf{Z}_0$, is  the characteristic impedance of plane waves in free space. It should be noted that the effect of periodic modulation results in the matrix representation of all circuit components (explained in the following text).
The tangential components of the total electric and magnetic fields on the surface are analogous to the total voltage $V_{\rm tot}$ and current $I_{\rm tot}$ in the circuit ($E_{\rm tot}^{t}\rightarrow V_{\rm tot}$ and 	$H_{\rm tot}^{ t}\rightarrow I_{\rm tot}$). 
From (\ref{Eq:field_expression}) we see that the voltage is composed of infinitely many harmonics:
	\begin{equation}
	V_{\rm tot}(z, t)=\sum_{n=-\infty}^{+\infty}v_n e^{-j(k_{zn} z-\omega_nt)}.\label{Eq: total voltage}
	\end{equation}
The total current is a superposition of all currents flowing in the conductance, inductance, and the shorted transmission line: $I_{\rm tot}=I_G+I_L+I_D$. The partial current in each circuit component is also a sum of an infinite number of Floquet harmonics,
	\begin{equation}
	I_{q}(z, t)=\sum_{n=-\infty}^{+\infty}i_{q,n} e^{-j(k_{zn} z-\omega_nt)}, \quad q\in[G,L,D]. \label{Eq: current composition}
	\end{equation}
Unlike the stationary scenario (without spatiotemporal modulation), in space-time modulated  systems,  $V_{\rm tot}(z, t)$ and $I_{q}(z, t)$ cannot be simply related by a scalar impedance or admittance.
Instead, one must consider all the harmonics from $n=-\infty$ to $n=+\infty$. Practically, we only take a finite number of harmonics from $n=-N$ to $n=+N$ into consideration ($N$ is large enough to ensure the convergence of harmonics).
The voltage and current at each element can be written as $2N+1$ dimensional vectors: $\accentset{\rightharpoonup}{v}=[v_{-N},v_{1-N},\cdots,v_N]^T$ and $\accentset{\rightharpoonup}{i}_q=[i_{q,-N},i_{q,1-N},\cdots,i_{q,N}]^T$. These voltage and current vectors can be related by an $(2N+1)\times (2N+1)$ admittance matrix $\mathbf{Y}_q$
	\begin{equation}
	\accentset{\rightharpoonup}{i_q} =\mathbf{Y}_q \cdot\accentset{\rightharpoonup}{v}. \label{hh}
	\end{equation}

Next, using the linear relations between voltage and current vectors, we find the admittance matrix for each lumped component in the circuit.
	The current flowing in the conductance reads
	\begin{equation}
	I_G(z, t)=G(z, t)V_{\rm tot}(z, t). \label{Eq: Ohm's_conductance}
	\end{equation}
After substituting Eqs.~(\ref{Eq: travelling_wave_conductance}), (\ref{Eq: total voltage}), and (\ref{Eq: current composition}) into Eq.~(\ref{Eq: Ohm's_conductance}), we have
	\begin{equation}
	\sum_{n=-\infty}^{+\infty}i_{G,n} e^{-j(k_{zn} z-\omega_nt)}=\sum_{\ell, m=-\infty}^{+\infty}g_mv_\ell e^{-j(k_{z,\ell+m} z-\omega_{\ell+m}t)} . \label{Eq:conductance_expansion}
	\end{equation}
Using the change of variable $\ell\rightarrow\ell-m$,  the $n$-th  current harmonic can be written as
	\begin{equation}
	i_{G,n}=\sum_{m=-\infty}^{+\infty}g_m v_{n-m}.\label{Eq: current_voltage_conductance}
	\end{equation}
We can see that the current harmonics are not only induced by the $n$-th voltage harmonic, but also contributed by coupling with other harmonics.
Equation~(\ref{Eq: current_voltage_conductance}) must be satisfied for all harmonics, generating a system of equations formed by  $2N+1$ linear equations ($n\in[-N,N]$). The equations can be written in the matrix form $\accentset{\rightharpoonup} {i_G}=\mathbf{Y}_G \accentset{\rightharpoonup} {v}$, which can be expanded as
	\begin{equation}
  \begin{pmatrix}
  i_{G,-N}  \\
  i_{G,1-N} \\
  \vdots   \\
  i_{G,N} \end{pmatrix}=\begin{pmatrix}
  g_{0} & g_{-1} & \cdots & g_{-2N} \\
  g_{1} & g_{0} & \cdots & g_{1-2N} \\
  \vdots  & \vdots  & \ddots & \vdots  \\
  g_{2N} & g_{2N-1} & \cdots & g_{0}
 \end{pmatrix}\begin{pmatrix}
  v_{-N}  \\
  v_{1-N} \\
  \vdots   \\
  v_{N} \end{pmatrix}.
	\end{equation}
Here, $\mathbf{Y}_G$ is called admittance matrix. 
It should be noted that $\mathbf{Y}_G$ is a Toeplitz matrix with elements $Y_G(s,t)=g_{s-t}$, where $s,t\in[-N,N]$ are the row and column indexes of the matrix, respectively.

Following a similar approach, we derive the admittance matrix for the space-time modulated inductance. The time-domain relation between the current and voltage across a time-varying inductor reads
	\begin{equation}
	I_{L}(z,t)L(z,t)=\int^t V_{\rm tot}(z,t^\prime) dt^\prime.\label{Eq:Ohm law}
	\end{equation}
After substituting Eq.~(\ref{Eq: travelling_wave_conductance}) and Eq.~(\ref{Eq: current composition}) into Eq.~(\ref{Eq:Ohm law}) and using the same mathematical treatments in Eq.~(\ref{Eq:conductance_expansion}), we can find the admittance matrix of inductance, $\mathbf{Y}_L$ which is filled by matrix element $Y_{ L}(s,t)=b_{s-t}/j\omega_t$ \cite{suppl}.

% the $n$-th harmonic of the current in the inductor can be written as
% \begin{equation}
% 	i_{L,n}=\sum_{m=-\infty}^{+\infty}\frac{b_m v_{n-m} }{j\omega_{n-m}}. \label{Eq: current_voltage_inductor}
% 	\end{equation}
% The system of equations defined by Eq.~(\ref{Eq: current_voltage_inductor}) can be written also in  the matrix form $\accentset{\rightharpoonup} {i_L}=\mathbf{Y}_L \accentset{\rightharpoonup} {v}$, where $\mathbf{Y}_L$ is the time-modulated inductor admittance matrix filled by elements $Y_{ L}(s,t)=b_{s-t}/j\omega_t$.

Finally, we need to calculate the admittance matrix of the shorted transmission line, $\mathbf{Y}_D$. One must notice that different Floquet modes have different frequencies and propagation constants  in the substrate. Since the substrate is not modulated, there are no coupling elements in $\mathbf{Y}_D$. Therefore,  $\mathbf{Y}_D$ is a diagonal matrix with elements $Y_D(n,n)=y_{D,n}$, where $y_{D,n}$ is the input admittance of shorted transmission line for $n$-th harmonic. Following the derivation in the textbook \cite[\textsection~2.7]{pozar2009microwave}, we can obtain the expression of $y_{D,n}$ \cite{suppl},  
	\begin{equation}
	y_{D,n}=\frac{1}{z_{D,n}^{TM}\tanh(jk_{xn}^{D} d)}, 
	\end{equation}
where $k_{xn}^{D}=\sqrt{\omega_n^2\epsilon_{\rm r}\epsilon_{\rm 0}\mu_0-k_{zn}^2}$ is the normal component of the wavenumber in the dielectric substrate and $z_{D,n}^{TM}=k_{xn}^{D}/\epsilon_{\rm r}\epsilon_{\rm 0}\omega_n$ is the TM-wave impedance of dielectric substrate.
%Considering harmonics from $-N$ to $+N$, $\mathbf{Y}_d$ is a diagonal matrix with elements $Y_d(l,l)=y_{dl}$, where $l\in[-N,+N]$.
	
After the admittance matrices for all the  circuit components are determined, the total admittance is defined as $\mathbf{Y}_{\rm tot}=\mathbf{Y}_G+\mathbf{Y}_L+\mathbf{Y}_D$. This matrix relates the total current and voltage (summation of the incident and reflected harmonics) of the circuit:
	\begin{equation}
	\accentset{\rightharpoonup} {i}^{\rm in}-\accentset{\rightharpoonup} {i}^{\rm re}=\mathbf{Y}_{\rm tot}\cdot ( \accentset{\rightharpoonup} {v}^{\rm in}+\accentset{\rightharpoonup} {v}^{\rm re}), \label{Eq: total field relation with admittance}
	\end{equation}
where the superscripts `in' and `re' stay for the incident and reflected harmonics, respectively. 
The tangential electrical and magnetic fields of the incident and scattered harmonics are related by the characteristic impedance matrix of free space $\mathbf{Z_{\rm 0}}$. For TM incidence, the impedance $\mathbf{Z_{\rm 0}}$ is a diagonal matrix filled with $Z_{\rm 0}(n,n)=z_{0,n}^{TM}={k_{xn}^{\rm 0}}/(\epsilon_0\omega_n)$, where $k_{xn}^{\rm 0}=\sqrt{\omega_n^2\epsilon_0\mu_0-k_{zn}^2}$ is the vertical wavevector of $n$-th harmonic. Therefore, we have
	\begin{equation}
	\accentset{\rightharpoonup} {v}^{\rm in}=\mathbf{Z}_{\rm 0}\cdot\accentset{\rightharpoonup} {i}^{\rm in}, \quad \accentset{\rightharpoonup} {v}^{\rm re}=\mathbf{Z}_{\rm 0}\cdot\accentset{\rightharpoonup} {i}^{\rm re} . \label{Eq: free space impedance}
	\end{equation}
	Substituting (\ref{Eq: free space impedance}) into (\ref{Eq: total field relation with admittance}),
	we can find $\accentset{\rightharpoonup} {i}^{\rm re}=-\mathbf \Gamma\cdot \accentset{\rightharpoonup} {i}^{\rm in}$. The parameter $\mathbf \Gamma$ is called the reflection matrix and is given by 
	\begin{equation}
	\mathbf \Gamma=(\mathbf{Y}_{\rm tot} \mathbf{Z}_{\rm 0}+\mathbf{I})^{-1}(\mathbf{Y}_{\rm tot} \mathbf{Z}_{\rm 0}-\mathbf{I}), \label{Eq: Reflectance}
	\end{equation}
where $\mathbf{I}$ is the $(2N+1)\times(2N+1)$ unity matrix.
The source vector is defined as $\accentset{\rightharpoonup} {i}^{\rm in}=[0,\cdots,0,1,0,\cdots,0]^T$. We  can see that the $n$-th reflected harmonic is equal to the matrix element located at the $n$-th row and $0$-th column of the reflection matrix $\mathbf \Gamma$, $i_n^{\rm re}= \Gamma (n,0)$. 

% The developed theory not only works for shunt-connected modulation components in Fig.~\ref{fig: circuit model}, but can be generalized to arbitrary connection forms, for example, series connection of $RLC$ components,  which will be discussed in Section~\ref{Sec: graphene}. The basic idea is to find the admittance matrix of a circuit branch relating the total voltage and current. The theory is applicable also to TE incidence. In fact, the polarization states only affect $\mathbf{Y}_D$ and $\mathbf{Z}_0$. The method can be easily adapted for TE-polarized wave after replacing $z_{D,n}^{TM}$ with $z_{D,n}^{TE}=\mu_0\omega_n/k_{xn}^D$, and $z_{0,n}^{TM}$ with $z_{0,n}^{TE}=\mu_0\omega_n/k_{xn}^{\rm 0}$.

If the surface conductance and inductance in Fig.~\ref{fig: circuit model} and the substrate properties are known, for a specific plane wave excitation the scattering harmonics are uniquely determined by Eq.~(\ref{Eq: Reflectance}). 
Alternatively, for desired scattering responses, it is also possible to find properly modulated conductance and inductance that correspond to the generation of the desired scattering harmonics by the metasurface. However,  there is no simple analytical formulation  that could directly determine the modulation functions for a set of desired scattering harmonics. For this reason, we mathematically optimize the Fourier coefficients of the modulation functions $G(z,t)$ and $B(z,t)$ until they  generate the wanted harmonics for two opposite incident directions ($\theta=+\theta_{\rm i}$ and $\theta=-\theta_{\rm i}$), realizing  desired nonreciprocal effects.

	\section{Engineering Nonreciprocity of Space-Time Modulated  Metasurfaces}\label{Sec: Functionalities}
	
In this section, we show that the generalized circuit model developed in Section~II can be used to realize  different nonreciprocal devices engineering  modulation functions.
As examples,  we design isolators, nonreciprocal phase shifters, quasi-isolators, and circulators on a physically rigid platform, only by modifying the modulation functions.

\subsection{Metasurface Isolators}\label{Sec: Isolators}
	
\begin{figure}[h!]
	\centering
	\includegraphics[width=0.8\linewidth]{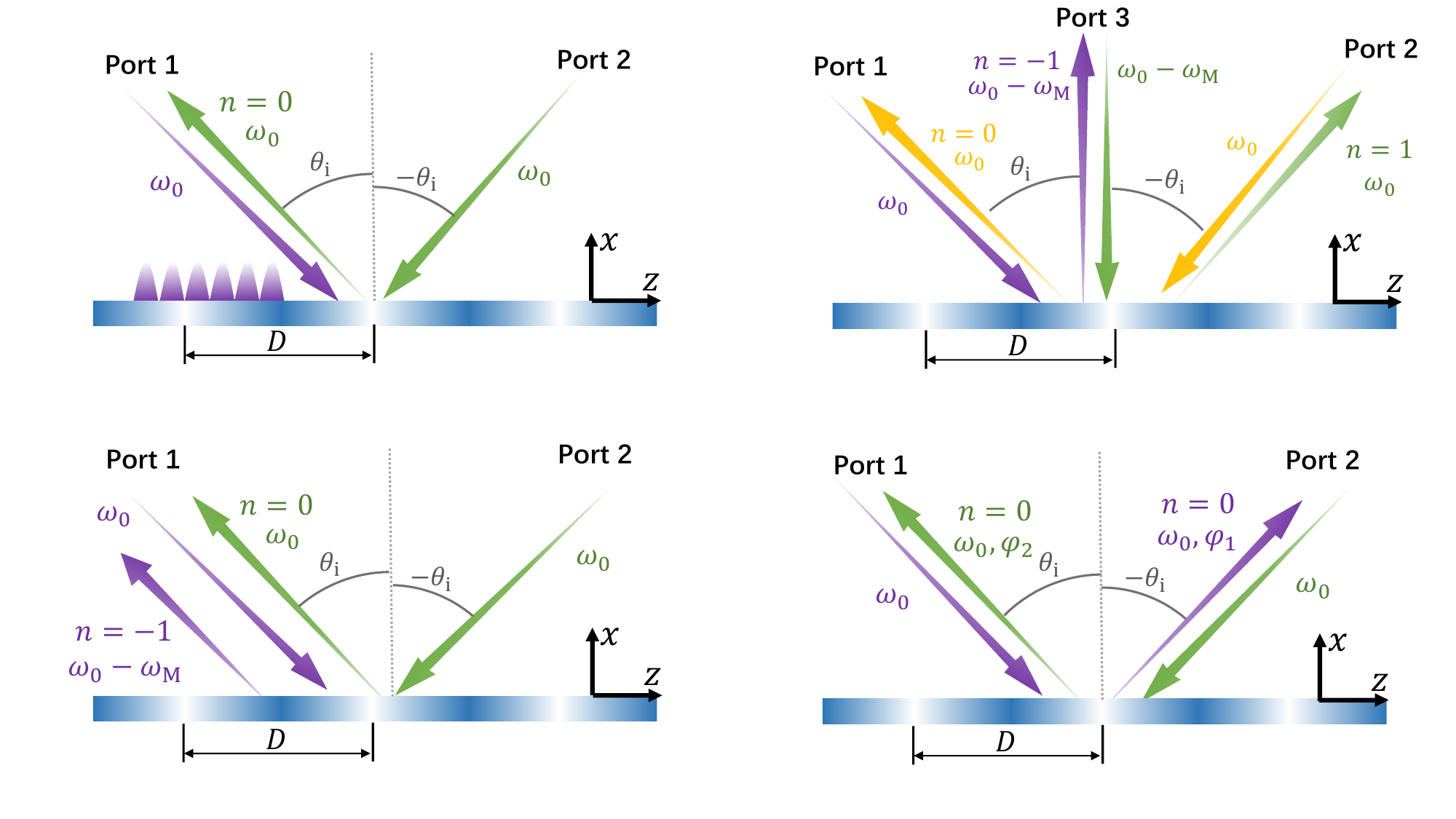}
	\caption{Schematic representation of the metasurface platform for implementing isolators. }\label{fig:isolator_lossy}
\end{figure}

An isolator is a lossy, passive, and matched two-port device that allows unidirectional wave transmission  \cite{pozar2009microwave}. 
For a perfect isolator, waves incident on one port are fully dissipated while the waves incident on the second port are fully transmitted without frequency conversion. The scattering matrix of an ideal isolator is defined as 
	\begin{equation}
	\overline{\overline S}= \begin{bmatrix}
	0       & 1  \\
	0       & 0
	\end{bmatrix}. \label{definition: isolator}
	\end{equation}
Figure~\ref{fig:isolator_lossy} shows the design principle of the metasurface isolator. Plane waves incident at $\theta=+\theta_{\rm i}$ (Port~1) are completely absorbed, while waves incident at  $\theta=-\theta_{\rm i}$ (Port~2) all are specularly reflected into Port~1 ($n=0$). It is important to notice that, according to this definition, there is no frequency conversion between ports.

Wave isolation realized with space-time modulated media has been demonstrated in a few papers \cite{shaltout2015time,chamanara2017optical,shi2017optical,correas2018magnetic}, where the classical scattering matrix [see Eq.~(\ref{definition: isolator})] is not suitable to describe those systems. In those works, the incident wave from the attenuation port is not absorbed due to material losses but converted to higher-order harmonics and transmitted, requiring absorptive filters to remove the generated harmonics. 
In another work \cite{hadad2015space}, although there is no frequency conversion of the scattered propagating waves, the waves incident from the attenuation port are not actually absorbed but leak away through other ports (transmission side).
Here, we propose a solution based on a lossy gradient surface impedance that allows us to design isolators according to the classical definition shown in  Eq.~(\ref{definition: isolator}).
%\textcolor{red}{The main action points are: (1) to create a perfect absorber with strong resonance, which can be realized by modulating a lossy surface impedance in space, and (2) to add temporal modulation on the gradient surface such that the resonances for different illumination directions can be spectrally resolved.}
The design methodology consists of three steps.

\underline{\textit{Step 1:}} Making use of the multi-channel metasurface concept, the first step in the design is to find the period of the metasurface space modulation that allows the required propagating channels. In this case, the two-port system will be implemented in the metasurface platform by the incident direction and the specular reflection, as it is shown in Fig.~\ref{fig:isolator_lossy}. This condition can be fulfilled by restricting the spatial period as $D<\lambda_0/(1+\sin\theta_{\rm i})$ \cite{suppl}. In addition, this periodicity ensures that both $S_{11}$ and $S_{22}$ are zero (no retroreflection or other higher order diffraction modes).

\underline{\textit{Step 2:}} Once the spatial period of the metasurface has been chosen, the second step  is to find a proper traveling-wave modulation profile that can realize perfect absorption for plane-wave incident from $\theta=+\theta_{\rm i}$ (Port 1).
We define the frequency of incidence as $\omega_0=\omega_{\rm d}$, and the modulation frequency is $\omega_{\rm M}=\omega_{\rm d}/10^3$. For total absorption at Port 1, the scattered field should have no specular reflection which means
\begin{equation}
 \left|\Gamma (0,0)\right|=0 \label{Eq: objectives_isolator_1}
\end{equation}
should be satisfied for incidence of $\theta=+\theta_{\rm i}$.
Next, we ensure strong \textit{evanescent fields} are excited simultaneously, which is the key point of this design method.
%With the defined periodicity,  the transverse wavenumber of propagating wave $k_0\sin\theta_{\rm i}$ is smaller than $\beta_{\rm M}$ (in this scenario), meaning that in one spatial period of incident fields, there exists multiple periods of surface impedance. For this reason, the reflected field is only affected by the average surface impedance during the modulation. However, the wavenumbers of evanescent waves $k_{zn}$ are always larger than $\beta_{\rm M}$. This indicates that multiple spatial periods of evanescent waves can exist on one period of surface impedance. 
The reason for the excitation of evanescent fields includes: (1) the wavelength of these fields are much smaller than the spatial modulation periodicity, and they are generated in order to locally satisfy the  gradient impedance boundary condition. Therefore, they tend to be sensitive to the surface boundary perturbation, as well as the incident angles; (2) the  amplitudes  of 
evanescent fields can control the resonance strength. To warrant the excitation of evanescent fields, we impose,
\begin{equation}
\left|\Gamma (1,0)\right|=A_1. \label{Eq: objectives_isolator_2}
\end{equation}
Notice that we only need to ensure the desired amplitude of the first-order evanescent mode ($n=1$).
Eq.~\ref{Eq: objectives_isolator_1} and Eq.~\ref{Eq: objectives_isolator_2} are called objective functions  that guarantee evanescent fields induced absorption at Port 1. 

Next we aim to find the optimal surface impedance that can satisfy these two objectives.
From the results of Section~\ref{Sec: Theory} we know that $ \Gamma (n,0)$ are determined by the Fourier coefficients of $G(z,t)$ and $B(z,t)$.   	In order to find $g_m$ and $b_m$ that satisfy the objectives, we can introduce unlimited numbers of $g_m$ and  $b_m$ in Eq.~(\ref{Eq: travelling_wave_conductance}). 
However, we assume that only three Fourier terms are enough to adequately  describe each  component [$g_{0,\pm 1}$ for $G(z,t)$ and $b_{0,\pm 1}$ for $B(z,t)$].
Without loosing generality, we also assume that $g_{m}=g_{-m}$ and $b_{m}=b_{-m}$ (this assumption always holds in the following text), such that the Fourier terms are real numbers and both $G(z,t)$ and $B(z,t)$ are even functions with respect to the  $z$-axis. 
Under these assumptions, the Fourier series for $G(z,t)$ and $B(z,t)$ in Eq.~(\ref{Eq: travelling_wave_conductance}) are reduced to
	%Considering each term is a complex number, therefore, we have four unknowns in total (real and imaginaray parts of $a_0$ and $a_1$)
	\begin{equation}
	G(z,t)=g_0+2g_1\cos(\beta_{\rm M}z-\omega_{\rm M}t)
	\end{equation}
	and 
	\begin{equation}
	B(z,t)=b_0+2b_1\cos(\beta_{\rm M}z-\omega_{\rm M}t)
	\end{equation}
At this point, in order to preserve the physical meaning of the model, one must ensure that the conductance and inductance are always positive  along the $z$-axis. To do that, two additional constraints are introduced, $g_0-2\left| g_1\right|>0$ and $b_0-2\left|b_1\right|>0$.
	
In total, there are two objective equations and two linear constraints to be considered with four unknown Fourier coefficients. %It is unlikely to analytically find the solutions due to the inverse operation of bulky matrix in (\ref{eq: scattered fields}).
To solve this optimization problem, we use MATLAB Optimization Toolbox to find the solutions of~(\ref{Eq: objectives_isolator_1}) and (\ref{Eq: objectives_isolator_2}), using the optimization function $f{mincons}$ (optimization code is available in \cite{suppl}). 
Table~\ref{tab:Table1} shows the optimized $b_m$ and $g_m$ for different evanescent mode amplitudes.

\begin{table}[tb]
\centering
 \caption{\label{tab:1}Optimized amplitudes of the Fourier harmonics for different values of the desired evanescent field amplitude. We assume $D=0.419 \lambda_{\rm d}$ ($\lambda_{\rm d}=2\pi c/\omega_{\rm d}$, where $c$ is the speed of light). The substrate has the relative permittivity  $\epsilon_{\rm r}=4$ and the thickness $d=0.133\lambda_{\rm d}$ (the same substrate is used throughout Section~\ref{Sec: Functionalities}).}
 \resizebox{0.48\textwidth}{!}{%
 \begin{tabular}{|c| c c| c c|c|c|} 
\toprule
 %&  \multicolumn{2}{|c|}{Conductance,  $G(z)$} &  \multicolumn{2}{|c|}{Inductance,  $B(z)=1/L(z)$} \\
\makecell{Evanescent field\\ $(n=1)$} & \makecell{ $g_0$ \\ $\times10^{-6}$ } & \makecell{ $g_1$ \\ $\times10^{-6}$} & \makecell{ $b_0$ \\ $\times10^{10}$ }& \makecell{ $b_1$ \\ $\times10^{10}$ } &\makecell{ $|S_{12}|^2$ \\ (dB) } & \makecell{ $|S_{21}|^2$ \\ (dB) } \\  \hline
\hline
 $A_1=1$& 208.15  & -38.1 & 43.36 & -11.42 &-21.6 &-83\\ \hline
 $A_1=3$& 26.17 & -5.50  & 36.03 & -3.46  &-5.37 &-64\\ \hline
$A_1=10$& 2.29  & -0.67  & 35.25 & -1.03  &-0.08 &-43.7\\ 
\bottomrule
 \end{tabular}}\label{tab:Table1}
\end{table}	
\begin{figure}[h!]
		\centering
		\includegraphics[width=1.0\linewidth]{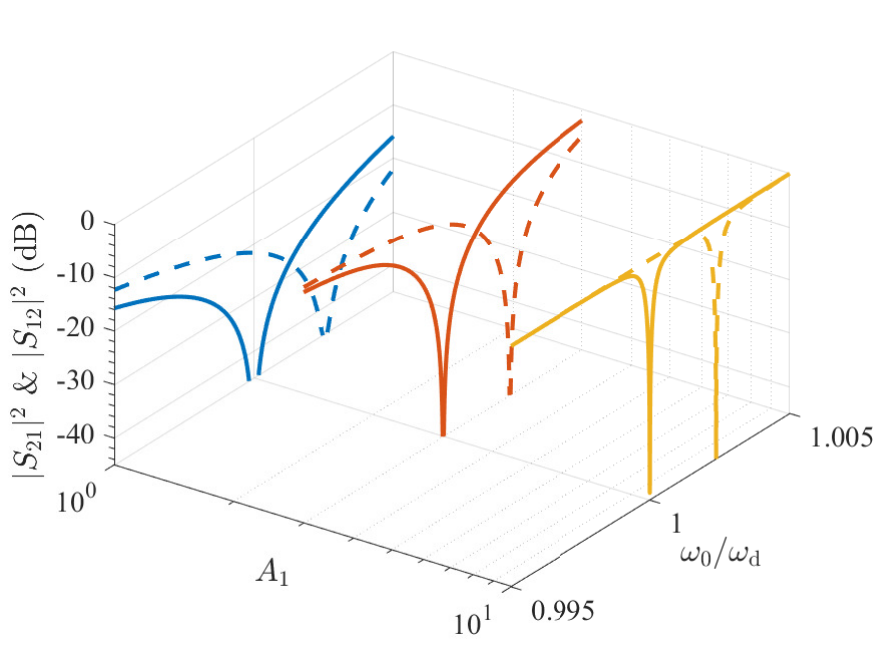}
		\caption{Transmittances for incidences from Port~1 ($S_{21}$: solid lines) and Port~2 ($S_{12}$: dashed lines). Here, $S _{21}=\Gamma(0,0)$ for incidence from Port~1, and also $S _{12}=\Gamma(0,0)$ for incidence from Port~2.  In the three scenarios, the designed amplitudes of $n=1$ evanescent mode  are  $A_1=1$ (blue), $A_1=3$ (red) and $A_1=10$ (yellow). }\label{fig:one_channel_spectrum_comparison}
	\end{figure}

\begin{figure}[h!]
\centering
\includegraphics[width=0.95\linewidth]{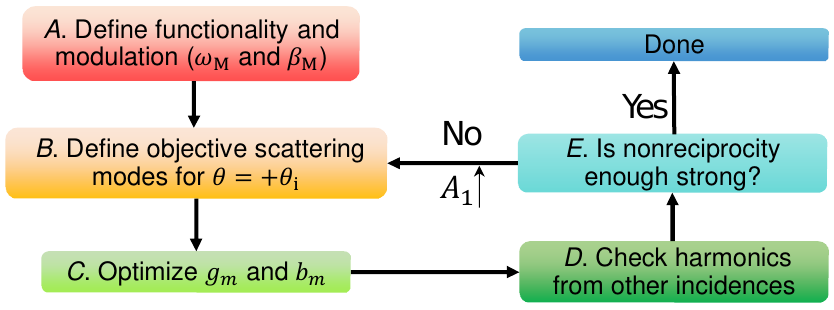}
\caption{Block diagram of the design procedure for all the exemplified devices proposed in Section~\ref{Sec: Functionalities}. }\label{fig:schematic_optimization}
\end{figure}

	\underline{\textit{Step 3:}} Once the surface impedance satisfying the objective functions at $\theta=+\theta_{\rm i}$ has been optimized, we examine its scattering fields for $\theta=-\theta_{\rm i}$ using the mode matching method presented in Section~\ref{Sec: Theory}. 
	%Let us first discuss the optimized surface defined with weak evanescent field $A_1=1$. 
	For incidence from Port 2, the resonance shifts to higher frequency (dashed line in Fig.~\ref{fig:one_channel_spectrum_comparison}). 
	The space-time modulated surface emulates a moving slab with velocity of $v_{\rm M}=\omega_{\rm M}/\beta_{\rm M}$ along the $+z$ direction. Effective motion of the metasurface compromises the established phase matching of the incident plane wave and one of the surface modes (as deigned at Step~2). For incidence from Port 2, the phase matching condition holds at lower frequencies. 
	This effect is similar to the phenomenon in \cite{hadad2015space}, based on asymmetric distortion of the dispersion diagram under travelling-wave modulation.
	For the case of $A_1=1$, although nonreciprocity is achieved, the isolation performance at the defined frequency $\omega_0=\omega_{\rm d}$ is not good, since the wave incident from port 2 is almost absorbed ($|S_{12}|^2<-20$~dB). 
	To obtain near perfect isolation, the amplitude of evanescent field $A_1$  should be further improved. We see that as $A_1$ increases, the resonance become stronger. When $A_1=10$ (yellow curves),  near-perfect isolation is achieved at $\omega_0=\omega_{\rm d}$.  In this case, wave incident from Port~1 is completely absorbed while from Port~2 it is fully delivered to Port~1  (the efficiencies of both absorption and transmission are above 99 \%).
The above design procedures are summarized in Fig.~\ref{fig:schematic_optimization}:  we first optimize the modulation functions that ensure the desired propagation modes as well as the evanescent modes for the forward direction (Steps A-C). Then, we examine the scattered harmonics for the backward direction and check the achieved level of nonreciprocity (Step D). If the nonreciprocity is not strong, we further enhance the amplitude of evanescent mode $A_1$ defined in Step B and optimize the modulation function until it satisfies the desired nonreciprocal functionality.  
	In the next subsections, based on this design procedure, we realize and engineer other nonreciprocal functionalities on the same physical platform. 
	
	\subsection{Metasurface Nonreciprocal Phase Shifters}
	
\begin{figure}[h]
		\centering
		\includegraphics[width=0.8\linewidth]{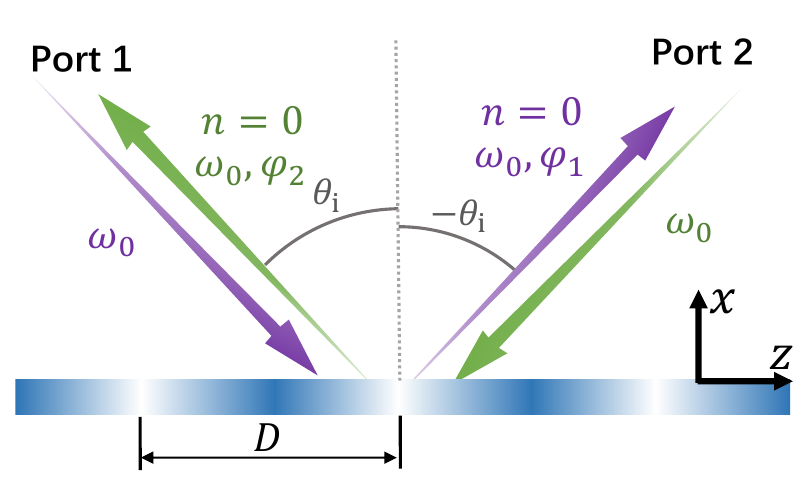}
		\caption{ Metasurface phase shifter for arbitrary phase conversion between two ports}\label{fig: phase_shifter}
\end{figure}

In lossless nonreciprocal phase shifters, the phase shifts for waves traveling between their two ports in opposite directions differ by a  specific phase difference  $\Delta\varphi$. The scattering matrix of a lossless and matched phase shifter can be generally represented as
\begin{equation}
	\overline{\overline S}= \begin{bmatrix}
	0        &e^{j(\varphi_1+\Delta\varphi)}  \\
	e^{j\varphi_1}       & 0
	\end{bmatrix}.
\end{equation}
The schematic of the proposed nonreciprocal metasurface phase shifter is shown in Fig.~\ref{fig: phase_shifter}. Light incident from $\theta=+\theta_{\rm i}$ (Port 1) is fully reflected (that is, transmitted into Port~2) with the transmission phase $\varphi_1$, while from the opposite direction $\theta=-\theta_{\rm i}$ (Port 2), the transmission phase is $\varphi_2=\varphi_1+\Delta\varphi$.
	
	%The metasurface is assumed to be lossless and only one scattering channel is open (specular). We specify the amplitude of first order evanescent fields for creating a resonance. Due to the lossless nature, the resonance will not show in the scattering magnitudes 
	
 As far as we know, space-time metasurfaces providing nonreciprocal phase shifts for free space propagation have not yet been reported. Here, we demonstrate that engineering induced evanescent fields it is possible to create metasurface phase shifters with controllable differential phases. The proposed design methodology consists of the following steps.

\underline{\textit{Step 1:}} Similarly to the design of metasurface isolators, the period of the metasurface phase shifter is selected  within the range  $D<\lambda_0/(1+\sin\theta_{\rm i})$ to allow only specular reflection. This condition will warrant $S_{11}=S_{22}=0$.

\begin{figure}[h]
	\centering
	\includegraphics[width=0.9\linewidth]{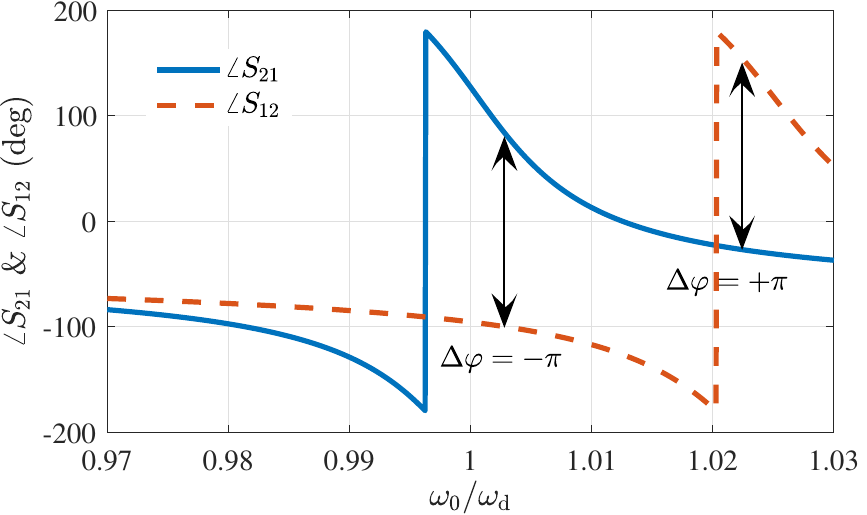}
	\caption{Transmission phases for waves incident at $\theta=+45^\circ$ (Port~1) and $\theta=-45^\circ$ (Port~2). The studied surface impedance is synthesized with $b_0=39.85\times10^{10}$ and $b_1=-7.50\times10^{10}$. The  spatial period  is $D=0.419 \lambda_{\rm d}$.}\label{fig:frequency_spectrum_phase_shifter}
\end{figure}

\underline{\textit{Step 2:}}  Next, we ensure proper excitation of resonant  evanescent fields for incidence of $\theta=+\theta_{\rm i}$. As we know from the design principle of the metasurface isolator (Section~\ref{Sec: Isolators}), such resonance is very sensitive to the transverse unidirectional modulation momentum and the resonance frequency is expected to shift at the opposite incidence ($\theta=-\theta_{\rm i}$).  Following this idea, we introduce one objective function $\left| \Gamma (1,0)\right|=A_1$ at $\theta=+\theta_{\rm i}$, which guarantees that the first order evanescent field is excited with amplitude $A_1$.
Since the system is required to be lossless,  the surface conductance is removed ($G=0$), and only the inductance is subject to modulation.
We assume a simple harmonic modulation waveform on $B(z,t)$
\begin{equation}
B(z,t)=b_0+2b_1\cos(\beta_{\rm M}z-\omega_{\rm M}t),
\end{equation}
where $\omega_{\rm M}=0.01\omega_{\rm d}$. Notice that the constraint $b_0-2\left|b_1\right|>0$ should be imposed to warrant that the inductance values remain positive. 
Using  optimization, a proper set of $b_0$ and $b_1$ can be found to make sure that $A_1=3$ (as an example).  We can see from Fig.~\ref{fig:frequency_spectrum_phase_shifter} (blue solid line) that the transmission phase has an abrupt change around the defined resonant frequency $\omega_{\rm d}$.

%Apparently, for a lossless and single-channel  metasurface (channel means scattering channel), the amplitudes of transmission from two complementary angles are always unity due to energy conservation. 
	
%However, this restriction is only imposed on transmission amplitudes.
%For the transmission phases, it is still  possible to be unequal in a time modulated system. 

	%Next, we demonstrate that the reflection phase conversion in space time modulated system can be realized by Doppler effect if strong evanescent modes exists on the pump direction. 

\underline{\textit{Step 3:}} After obtaining the optimal surface impedance, we investigate its transmission phase for the opposite incidence. Similar to the case of metasurface isolators (see Section~\ref{Sec: Isolators}),  sharp resonance in the phase spectrum shifts to higher frequency, creating a large  phase difference at a fixed frequency. For example, at $\omega_0=1.003\omega_{\rm d}$ and $\omega_0=1.023\omega_{\rm d}$,  the devices exhibits $-\pi$ and $+\pi$ phase shifts for waves traveling in the opposite directions, which can be practically used as a gyrator. 
The phase shift is continuously tunable by increasing or decreasing the modulation frequency, which adjusts spectral distances of the two resonances.
The frequency bandwidth of such phase shifters depends on the quality factor of the resonance which is determined by $A_1$. Decreasing  $A_1$ and  increasing the modulation frequency we can obtain a more flat phase curve which effectively expands the frequency bandwidth.

	\subsection{Metasurface Quasi-Isolators}
	
\begin{figure}[h]
		\centering
		\includegraphics[width=0.8\linewidth]{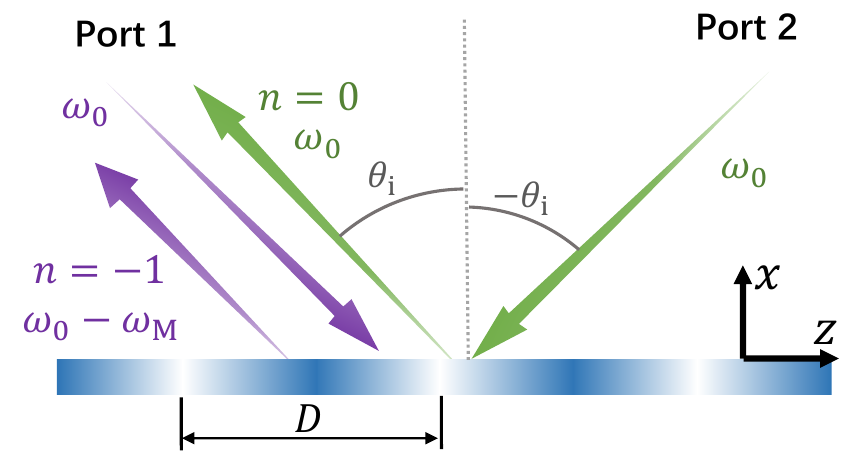}
		\caption{ Schematic represenation of a  quasi-isolator implemented in the proposed metasurface platform. }\label{fig:isolator_lossless}
	\end{figure}

In Section~\ref{Sec: Isolators}, we have shown how one can realize wave isolators  using lossy modulated impedance sheets  and demonstrated near perfect isolation without the need of frequency filters.
Here, we show that wave isolation can be realized also in lossless two-port systems, as schematized in Fig.~\ref{fig:isolator_lossless}.
Instead of being absorbed, as in usual isolators, the energy from Port~1 is fully reflected black (retro-reflection) with a frequency conversion $\omega_0\rightarrow \omega_0-\omega_{\rm M}$, while the energy from Port~2 is fully delivered to Port~1 (specular reflection) without frequecny conversion $\omega_0\rightarrow \omega_0$.  It is important to notice that, although frequency conversion occurs when the system is illuminated from Port~1 ,  this higher-order harmonic is not delivered to Port~2 and, for this reason, frequency filters are still not necessary. 
These devices provide the same functionality as ideal isolators if one considers the signal frequency, and therefore we call them as quasi-isolators. Note that ideal lossless isolators without frequency conversion or conversion to some other mode or channel are forbidden by energy conservation. The design of these devices has two steps.

\underline{\textit{Step 1:}} The first step is to define the spatial period of the metasurface that allows not only specular reflection but also retro-reflection. This requirement can be satisfied choosing  $D=\lambda_0/(2\sin\theta_{\rm i})$. In this case, we still create a two-port system, but the incident waves are allowed to reflect back into the same port (retro-reflection). Notice that the difference with the lossy isolator where the condition $S_{11}=S_{22}=0$ was automatically fulfilled by choosing the period which does not allow propagation of any diffracted mode, including the retro-reflected one.  

\underline{\textit{Step 2:}}	The second step is to realize perfect retro-reflection for incidence from Port~1 ($\theta=+\theta_{\rm i}$) at the target frequency $\omega_{\rm d}$.  Notice that in the case of ideal retro-reflection the energy transmitted to Port~2 is zero due to energy conservation. 
%We enforce the occurrence of full retroreflection at $\omega_0=\omega_{\rm d}$ when incident from $\theta_0=+\theta_{\rm i}$.
In addition, the evanescent fields should be specified to control the $Q$-factor of the system. 
Taking these factors into considerations, we define two objectives for incidence from $\theta=+\theta_{\rm i}$:
	\begin{equation}
	\left|\Gamma (-1,0)\right|=1,  \quad \left|\Gamma (1,0)\right|=A_1, \label{Eq:objective function 2}
	\end{equation}
Combining these objectives with the linear condition that warrants only positive values of the inductance, $B(z,t)>0$, one can see that at least three degrees of freedom are required for specifying the system. For this reason, three unknowns ($b_0$, $b_{1}$, and $b_{2}$) are introduced to synthesize   $B(z, t)$. We can write the expression for $B(z, t)$ as
\begin{equation}
	B(z,t)=b_0+\sum_{m=1}^{m=2}2b_m\cos[m(\beta_{\rm M}z-\omega_{\rm M}t)].
\end{equation}
where modulation frequency $\omega_{\rm M}=2\omega_{\rm d}/10^4$ is assumed in this example. 
Using optimization tools, we find the Fourier coefficients for three different values of $A_1$, shown in Table~\ref{tab:Table2}.

	\begin{table}[h]
\centering
 \caption{Optimized amplitudes of the Fourier harmonics for different values of the evanescent-field amplitude. }
 \resizebox{0.41\textwidth}{!}{%
 \begin{tabular}{|c| c c c|c|c|} 
\toprule
 %&  \multicolumn{2}{|c|}{Conductance,  $G(z)$} &  \multicolumn{2}{|c|}{Inductance,  $B(z)=1/L(z)$} \\
\makecell{Evanescent field\\ $(n=1)$} &  \makecell{ $b_0$ \\ $\times10^{10}$ }& \makecell{ $b_1$ \\ $\times10^{10}$ }&
\makecell{ $b_2$ \\ $\times10^{10}$}&\makecell{ $|S_{12}|^2$ \\ (dB) }&\makecell{ $|S_{21}|^2$ \\ (dB) } \\  \hline
\hline
 $A_1=1$& 229.83  & 30.19 & -37.70 & -15.73 &-81.87\\ \hline
 $A_1=3$& 192.52 & 12.34  & -13.86 &-2.41 &-46.02\\ \hline
$A_1=10$& 187.79  & -6.30  &6.13 &-0.04 &-42.42\\
\bottomrule
 \end{tabular}}\label{tab:Table2}
\end{table}	
	
	\begin{figure}[h!]
		\centering
		\includegraphics[width=0.8\linewidth]{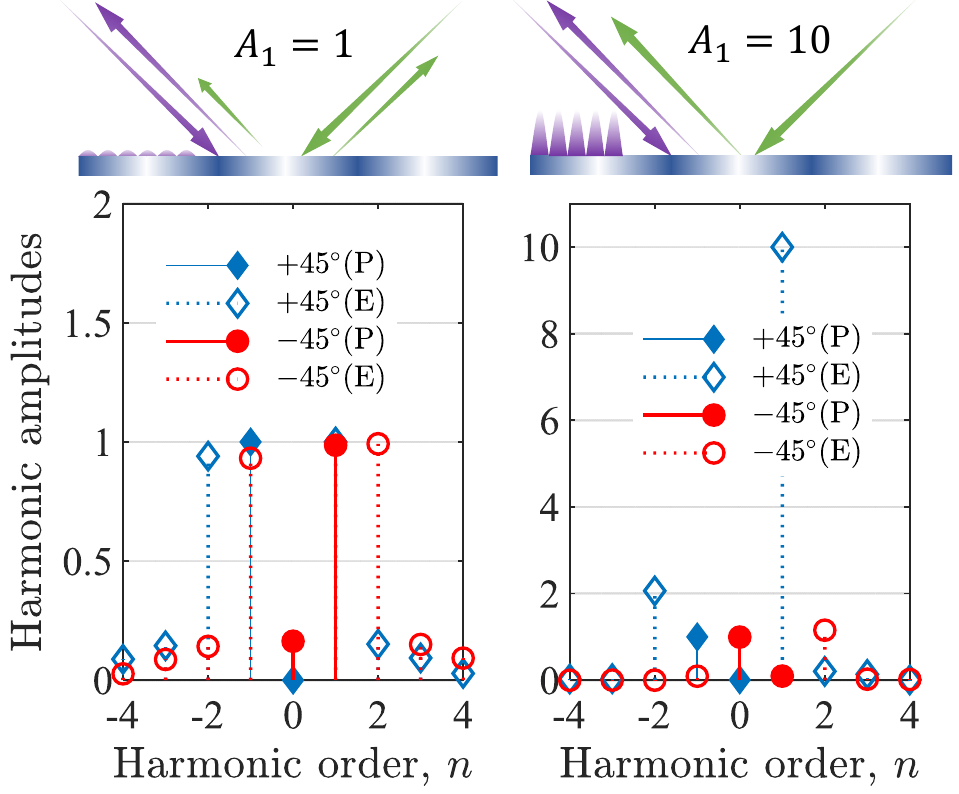}
		\caption{The amplitudes of excited harmonics for incidences from $+45^\circ$ and $-45^\circ$ at $\omega_0=\omega_{\rm d}$. From left to right panels, the amplitudes of evanescent mode $n=+1$ are $A_1=1$ and $A_1=10$, respectively. In the legends, \textquoteleft P\textquoteright~means propagating modes and \textquoteleft E\textquoteright~means evanescent modes. }\label{fig:two_channel_harmonic_comparison}
	\end{figure}

Figure~\ref{fig:two_channel_harmonic_comparison} presents the scattered harmonics at two opposite incidence for two optimal surfaces.
The first panel shows the results for $A_1=1$. We can see that the harmonics excited by plane waves at Ports~1 and~2 (incidence angles $+45^\circ$ and $-45^\circ$) are almost identical, meaning that if the amplitude of the first-order  evanescent harmonic is so small, there is no significant phase mismatch for excitation from the other port. Therefore,  waves incident on both ports are reflected back into the same port. However, as $A_1$ increases, the enhanced resonance of evanescent fields becomes more sensitive to the  angle of illumination. For the case  when $A_1=10$, the excited evanescent modes are completely different for  illuminations from the opposite angles, which leads to  enhanced transmission from Port~2 to Port~1 and realization of the quasi-isolator regime. 
%Considering the low pumping frequency, which is a common restriction in practical modulation platform, the above studies have practical meaning.  
These results again show that we can impose strong evanescent fields (improve $Q$-factor) to achieve evident efficiency of the nonreciprocal devices.

	%We expected that for incidence from the opposite direction, the excited evanescent modes would be different with the forward incidence due to phase mismatch. The induced current on the surface impedance is also different, thus its reradiated fields will not directed to retro direction.
	
\subsection{Metasurface Circulators}
	
\begin{figure}[h]
		\centering
		\includegraphics[width=0.8\linewidth]{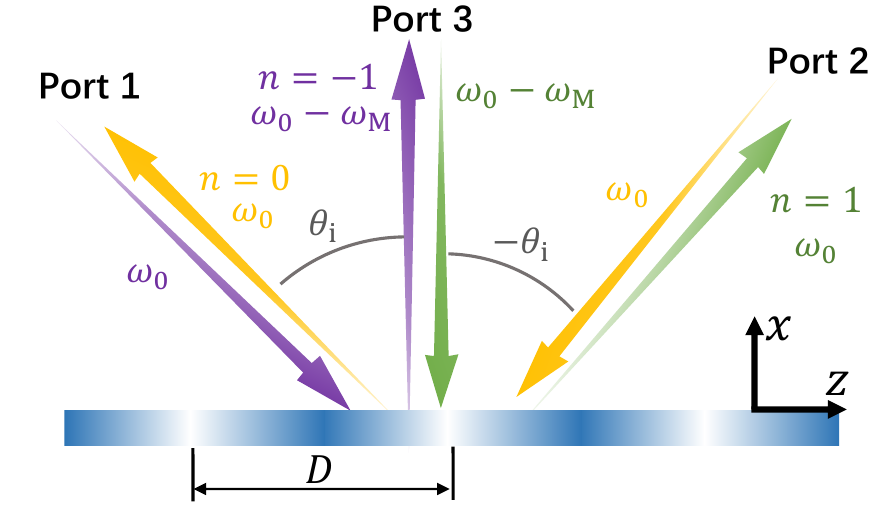}
		\caption{ Schematic representation of a metasurface circulator. }\label{fig:circulator}
	\end{figure}

With the proper choice of spatial period and using more complex time-modulation function, it is possible to control nonreciprocal wave propagation between more ports. As a particular example, we use the  design method explained above to realize  three-ports metasurface circulators.  Figure~\ref{fig:circulator} shows the schematics of the proposed metasurface circulator, where the incident wave travels unidirectionally between three ports: 1~$\rightarrow$~3,  3~$\rightarrow$~2, and 2~$\rightarrow$~1. It is important to note that the device introduces frequency down-conversion ($\omega_0\rightarrow\omega_0-\omega_{\rm M}$)  for propagation from 1~$\rightarrow$~3, and up-conversion ($\omega_0-\omega_{\rm M}\rightarrow\omega_0$) for propagation from  3~$\rightarrow$~2.
A similar functionality has been realized in a recent work \cite{shi2017optical} exciting two independent optical modes with transition momentum ($\beta_{\rm M}$, $\omega_{\rm M}$) in the dispersion diagrams and creating a high-quality resonator system. 
However, the known design method needs careful numerical optimizations of the band structures to ensure proper support of multiple optical modes. In addition, these optical modes should be tailored to be strongly resonant (high $Q$), to spectrally resolve them. 
	
Here, we show that using the mode-matching method and mathematical optimization  one can easily find surface impedances that realize perfect wave circulation, thus avoiding complicated band structure engineering. The design methodology for metasurface circulators consists of  two steps.
	
\underline{\textit{Step 1:}} First,	we increase the spatial period of the  metasurface to $D=\lambda_0/\sin{\theta_{\rm i}}$. This periodicity allows propagation of two additional diffracted modes for the incidence angle of $\theta=\theta_{\rm i}$ (wave from Port 1 can be retro-reflected or delivered to Port 3), so that the metasurface can be viewed as a three-port system (see Figure~\ref{fig:circulator}).  

\underline{\textit{Step 2:}} The second step is to find the modulation functions for the metasurface circulator design.
Enforcing that waves incident from Port~1 are fully directed to Port~3 and considering that the system is space-time modulated ($\omega_{\rm M}\neq0$), the amplitude of the $n=-1$ diffraction mode should be $\left|\Gamma (-1,0)\right|=1/\cos{\theta_{\rm i}}$ due to the energy conservation principle \cite{asadchy2016perfect}.
This condition automatically ensures photonic transition between two different optical modes as \cite{shi2017optical} aims to achieve. Then we prescribe the amplitude of the first-order evanescent field harmonic as $\left| \Gamma (1,0)\right|=A_1$, to control the quality factor of the system. 
	
\begin{figure}[h!]
		\centering
		\includegraphics[width=1\linewidth]{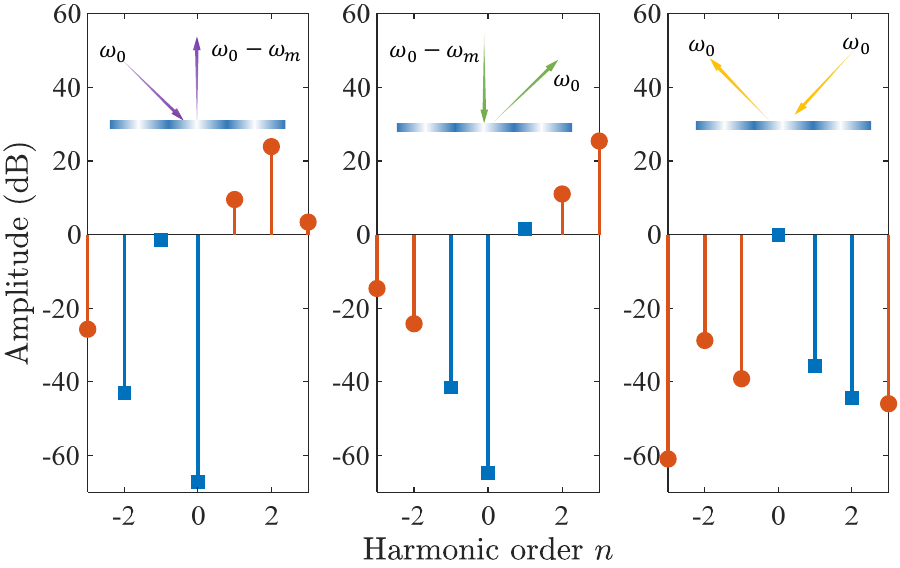}
		\caption{ Calculated harmonics for excitation from (a) $\theta=+45^\circ$, (b) $\theta=0^\circ$ and (c) $\theta=-45^\circ$.  The modulation frequency is $\omega_{\rm M}=\omega_{\rm d}/10^3$. The blue squares represents for propagating modes and the orange circles means evanescent modes. The optimized harmonic amplitudes of $B(z,t)$ are $b_0=198.54\times10^{10}$, $b_1=22.71\times10^{10}$, $b_2=8.77\times10^{10}$, $b_3=7.18\times10^{10}$ and $b_4=1.47\times10^{10}$. The prescribed evanescent field amplitude is $A_1=3$. }\label{fig:harmonic_spectrum_circulator}
	\end{figure}

In the system optimization we define two objectives corresponding to the above requirements  for harmonics. However, we should notice that, as the number of ports increases, the positive-inductance condition is more difficult to satisfy if we only introduce a few Fourier terms in the modulation function.
For this reason, more Fourier harmonics are included in $B(z,t)$ to provide more freedoms in the construction of a purely inductive modulated sheet impedance. In particular, we introduce five unknowns in $B(z,t)$:
	\begin{equation}
	B(z,t)=b_0+\sum_{m=1}^{m=4}2b_m\cos[m(\beta_{\rm M}z-\omega_{\rm M}t)]
	\end{equation}
Next, the Fourier harmonics of $B(z, t)$  are optimized, and the resulting  values for the harmonics are summarized in the caption of Fig.~\ref{fig:harmonic_spectrum_circulator}. 
	
Now, we examine the scattering harmonics when the circulator is illuminated from each of the three ports. The results are shown in Fig.~\ref{fig:harmonic_spectrum_circulator}. 
In the first scenario (left panel in Fig.~\ref{fig:harmonic_spectrum_circulator}), the wave incident from $\theta=+\theta_{\rm i}$ is fully delivered to the normal direction (Port~3) with strong evanescent fields excited on the surface. In the second scenario (middle panel in Fig.~\ref{fig:harmonic_spectrum_circulator}), the wave incident from Port~3 goes to Port~2  and excites the same sets of evanescent fields with scenario~1. This behavior is ensured by the phase matching condition established for scenario~1.
	%the  perfectly satisfy the modulating boundary condition. at this point, the incident wave, reflected wave and evanescent wave When incident from $P_2$, the energy is channeled to $P_3$
	%The circulation is realized with frequency conversion during $P_1\rightarrow P_2$ and $P_2\rightarrow P_3$, due to the restrictions from Floquet theorem.
Finally, for the wave incident from $\theta=-\theta_{\rm i}$ (Port~3), the incident wave propagates in the opposite direction with the modulation wave. Due to the strong phase mismatch, the resonance disappears and the device behaves as a normal mirror with full transmission to Port~1.
	
As the number of channels increases, more Fourier harmonics should be incorporated in the design of the modulation function, which opens up the possibility to realize wave circulation between five or more ports.

\newenvironment{thisnote}{\par\color{red}}{\par}

\section{Possible Implementations in microwave and terahertz ranges} \label{Sec: implementation}

% In this section, we discuss possible implementation platforms for different frequency ranges. The key point is to find a tuanble structure that can modify surface impedance First, we propose the use of Intelligent Hardware platforms for the implementation at microwave and mmWave. 
% First, we discuss the use of graphene as mechanism to tune the surface impedance at terahertz frequencies. 

In Section \ref{Sec: Functionalities} we have demonstrated that various nonreciprocal functionalities can be realized within the proposed space-time modulated surface impedance platform. 
Surface impedance is one of the most commonly used models for the design of metasurfaces, see e.g.  \cite[\textsection~2.4.3]{yang_rahmat-samii_2019}.
Required surface impedance can be implemented by structuring a metallic layer into periodic patterns \cite{wang2018systematic,wang2018extreme}. Dynamical modulation of surface impedance can be realized by embedding tunable elements or materials in meta-atoms (unit cells of the metasurface), e.g.  \cite{li2018metasurfaces}.
% For a spatially gradient surface impedance, it is necessary to discrete it and then locally implement it  with real materials \cite{wang2018extreme}. 
In microwave and millimeter-wave bands, surface impedance of meta-atom arrays can be controlled by varactors and varistors. This method has been experimentally demonstrated in the design of tunable reciprocal metasurfaces \cite{chicherin2006mems,sievenpiper2003two,sievenpiper2002beam}, as well as nonreciprocal space-time devices by modulating varactors in a travelling-wave form \cite{hadad2016breaking,zang2019nonreciprocal,taravati2017nonreciprocal}.
In the terahertz range, tunable lumped components are not practical to use due to the reduced footprint. Graphene, a two-dimensional material with gate-controlled surface impedance, can be used for realization  of spatiotemporal surface impedances due to its gate-controlled sheet  impedance. 
In this section, we discuss implementation possibilities of the proposed multifunctional platform, based on tunable lumped components in the  microwave band and graphene in the terahertz band.

\subsection{Implementation Based on Reconfigurable Intelligent Surfaces}

% In section \ref{Sec: Functionalities} we have demonstrated that various nonreciprocal functionalities can be realized within the proposed space-time modulated surface impedance platform. 
% Surface impedance is one of the most commonly used model for the design of metasurfaces\cite[\textsection~2.4.3]{yang_rahmat-samii_2019}.
% A rigid surface impedance can be implemented by structuring a metallic layer into periodic patterns \cite{wang2018systematic}.
% For a spatially gradient surface impedance, it is necessary to discrete it and then locally implement it  with real materials \cite{wang2018extreme}. 
% Surface impedance of meta-atoms (unit cell of metasurface) can also be controlled by varactors and varistors, providing dynamical tunability of the metasurfaces.  
% This method has been experimentally demonstrated in the design of tunable reciprocal metasurfaces \cite{chicherin2006mems,sievenpiper2003two,sievenpiper2002beam}, as well as nonreciprocal space-time devices by modulating varactors in a travelling-wave form \cite{hadad2016breaking,zang2019nonreciprocal,taravati2017nonreciprocal}. 

With modern integrated-circuit technology, both  varactors and varistors can be packaged inside a single chip and their values can be independently controlled by external signals \cite{liu2019electromagnetic}. 
The control chip is then connected to each meta-atom to dynamically modify the local surface resistance and reactance of metasurfaces. 
The schematic of a suitable hardware platform has been proposed in  \cite{pitilakis2018software,tasolamprou2019exploration}. Meta-atoms are formed by four patches and each of them is pinned to the other side of metasurfaces and connected to controller chips (see Fig.~1 in \cite{pitilakis2018software}). The ground plane is positioned between the patch layer and the chip layer, which reduces  interactions between electromagnetic waves and control circuitry. 
The  chips are connected with each other by signal lines. With proper dictations from the computer,  the meta-atoms are properly modulated   \cite{petrou2018asynchronous}, to control the effective surface impedance of each element. 
Such modulation platform has been utilized to achieve reconfigurability of reciprocal metasurfaces \cite{liu2019intelligent} where the electrical properties of the surface is modulated in space.
By applying a time-varying control signal upon each chip, the surface impedance of the metasurface can be spatiotemporally modulated and the modulation function can change in time to provide multiple nonreciprocal functionalities as required.
% as we
% our proposed theoretical model for space-time multifunctional metasurface can also be realized via this platform 

It is worth to mention that in the theoretical model presented here we assume the modulation of inductance, but in practical implementation it is usually easier to realize temporal modulation of capacitance than inductance. The design equations for capacitive reactance can be derived in the same way as shown above for inductive sheets. Alternatively, it may be reasonable to mimic time-varying inductance using a large fixed inductance in series with a time-modulated capacitance of the control chip. In this way, the total reactance is inductive and modulated in time, producing the same effect as a time-varying inductance.

% For actual implementations, the modulations of surface impedance can be achieved in many means. In the microwave band,   integrated circuits (ICs) with incorporated  varactors and varistors can be embedded into the metasurface, in order to independently control the local resistance and reactance \cite{liu2019intelligent} of the meta-atoms. In this case, it is possible to implement all the  functionalities discussed in Section \ref{Sec: Functionalities}, and more. 
% It is worth to mention that in practice it is usually easier to realize temporal modulation of capacitance than inductance. For this reason, it may be reasonable to mimic time-varying inductance using a large fixed inductance in series with a time-modulated capacitance. In this way, the total reactance is inductive and modulated in time, producing the same effect as time-varying inductance.

\subsection{Implementations Based on  Graphene Platform } \label{Sec: graphene}
	
In the terahertz and mid-infrared band, graphene is a good candidate to implement the tunable component of space-time modulated metasurfaces due to its tunable electrical conductivity and the compatibility with the advanced nanofabrication technologies \cite{salary2018time,correas2018magnetic,qin2016nonreciprocal,correas2016nonreciprocal, salary2018electrically}.
	Considering that graphene is a lossy material, in this section we show how one can use space-time modulated graphene sheets to create  tunable wave isolators in the terahertz range. 
	
	It is known that the sheet conductivity of graphene can be effectively modified via electrical bias or optical pumping. This provides the possibility of synthesizing space-time modulated surface impedances by locally varying the bias voltage applied on graphene. The sheet conductivity of graphene in the low terahertz range can be modeled as
	\begin{equation}
	\sigma_{\rm s}=\frac{\sigma_0}{1+j\omega\tau},\label{Drude}
	\end{equation}
	where 
	\begin{equation}
	\sigma_{\rm 0}=\frac{e^2\tau k_{\rm B }T}{\pi \hbar^2}\left[\frac{E_{\rm F0}}{k_{\rm B }T}+2\ln\left(e^{\frac{-E_{\rm F0}}{k_{\rm B }T}}+1\right)\right] \label{DC}
	\end{equation}
	is the DC conductivity, $\tau$ is the electron scattering time, $e$ is the electron charge, and $E_{\rm F0}$ is the Fermi level of graphene which can be controlled by the external voltage. As can be seen from  Eqs.~(\ref{Drude}) and (\ref{DC}), graphene is naturally resistive and inductive in the terahertz range,  and its sheet impedance can be expressed as $Z_{\rm g}=1/\sigma_{\rm  s}=R_{\rm s}+jX_{\rm s}$. If spatiotemporally varying gate voltage is exerted on a graphene strip, both of its sheet resistance ($R_{\rm s}=1/\sigma_{\rm 0}$) and inductance ($L_{\rm s}=\tau/\sigma_{\rm 0}$) can be harmonically modulated  with the same pumping scheme.
	\begin{figure}[h!]
		\centering
		\includegraphics[width=0.95\linewidth]{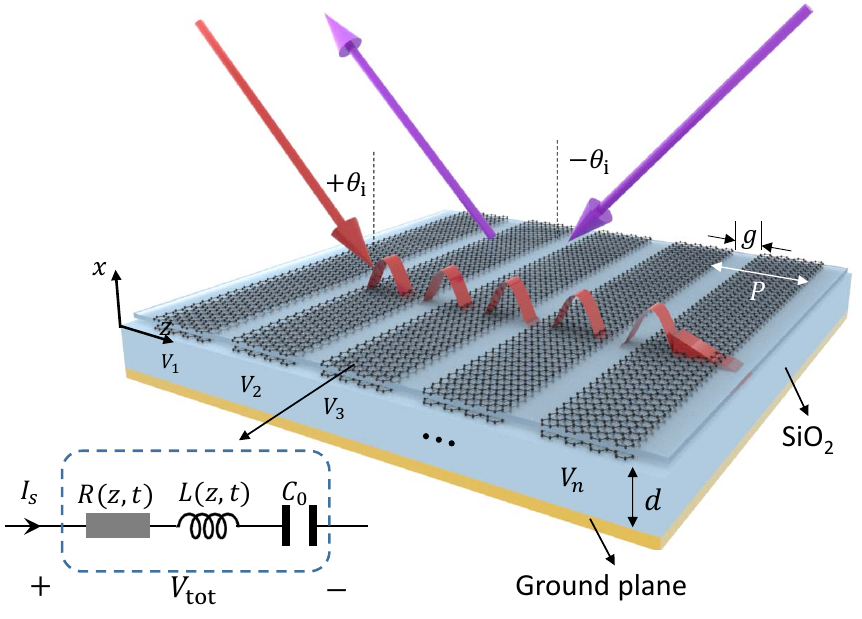}
		\caption{Schematic representation of a practical design. The surface impedance of one unit cell is implemented by  periodically  arranged graphene strips (sub-cells).  
			The size of one sub-cell is $P=2~\mu$m, and the gap width between two neighboring strips is $g=100$~nm. The permittivity of  ${\rm SiO_2}$ substrate  is $\epsilon_{\rm r}=4$, and its thickness equals $d=4~\mu$m.
			The local gate voltages on graphene are changing with time.
			When a wave is incident from $\theta=+\theta_{\rm i}$, it is transformed into evanescent fields and eventually dissipated in graphene (red wave). For the incidence at $\theta=-\theta_{\rm i}$, it is reflected in the specular direction (purple wave). }\label{fig:Graphene_model}
	\end{figure}
	
	Figure \ref{fig:Graphene_model} shows the schematic representation of the proposed design where graphene strips are periodically positioned on top of a metallized substrate. For the convenience of gating, a pair of graphene strips are  separated by a thin dielectric film (e.g., ${\rm Al_2O_3}$) to form a self capacitor \cite{correas2018magnetic}. Due to the stacked structure, the effective sheet impedance of graphene is reduced to half of the single-layer graphene impedance.  Under TM polarized incidence, the electrical response of graphene patterns can be described as a series of resistance, inductance and capacitance (all of them are effective values). 
	The static resistance and inductance of graphene without modulation are 
	\begin{equation}
	R_0=\frac{R_{\rm s}}{2}\left(\frac{P-g}{g}\right), \quad L_0=\frac{L_{\rm s}}{2}\left(\frac{P-g}{g}\right),
	\end{equation}  
	where $P$ is the period of the graphene strip array, $g$ is the distance between two adjacent graphene strips, and $(P-g)/g$ is the structural factor which takes into account the enlargement of  resistance and inductance due to the patterning of graphene \cite{wang2018systematic}. 
	The capacitance $C_0$ is contributed by capacitive coupling of the adjacent strips \cite{luukkonen2008simple}, 
	\begin{equation}
	C_0=\frac{2}{\pi}\epsilon_{\rm eff}\epsilon_0 P\ln\left[\csc\left(\frac{\pi g}{2P}\right)\right] \label{eq: capacitance_graphene}
	\end{equation}
	with the effective permittivity $\epsilon_{\rm eff}=(\epsilon_{\rm r}+1)/2$. Notice that Eq.~(\ref{eq: capacitance_graphene}) becomes invalid for very thin substrates ($d<0.3P$), because in this case the near-field coupling between graphene strips  and the ground plane should be also  considered \cite{tretyakov2003dynamic}. 
	In the present design ($d>0.3P$), Eq.~(\ref{eq: capacitance_graphene}) is valid and the capacitance is determined only by the structuring of graphene and the substrate permittivity, therefore it does not change during the modulation.
	A pump voltage is exerted on graphene, and the resistance and inductance of graphene are modulated around their static values. We assume the modulation function in form
	\begin{equation}
	f_{\rm M}(z,t)=a_0+\sum_{\substack{m=1}}^{+\infty}2a_m\cos[m(\beta_{\rm M}z-\omega_{\rm M}t)] \label{modulation function_graphene}
	\end{equation}
	with $a_0=1$. Thus, the space-time lumped values are $R(z,t)=R_0 f_{\rm M}(z,t)$ and $L(z,t)=L_0 f_{\rm M}(z,t)$.
	
	The current $I_S(z,t)$ and voltage $V_{\rm tot}(z,t)$ of the series circuit are related as
	\begin{equation}
	\begin{split}
	V_{\rm tot}(z,t)&=R(z,t)I_S(z,t)+L(z,t)\frac{dI_S(z,t)}{dt}\\
	&+I_S(z,t)\frac{dL(z,t)}{dt}+C_0^{-1}\int^t I_S(z,t^\prime) dt^\prime . \label{2}
	\end{split}
	\end{equation} 
	Following the same mathematical treatment as in Section \ref{Sec: Theory}, we obtain the coupling equation
	\begin{equation}
	\sum_{m=-\infty}^{+\infty}\left( R_0+j\omega_n L_0\right)a_m i_{S,n-m}
	+\frac{1}{j\omega_nC_0}i_{S,n}=v_n ,
	\label{Eq:boundary condition}
	\end{equation} 
	where  $i_{S,n}$ and $v_n$ are the $n$th harmonic components of $I_S$ and $V_{\rm tot}$, respectively. The voltage and current vectors are related by the impedance matrix, $\mathbf{Z}_S \accentset{\rightharpoonup}{i}_S=\accentset{\rightharpoonup}{v}$.
 $\mathbf{Z}_S$ is a $(2N+1)\times(2N+1)$  matrix filled with elements
	\begin{equation}
	Z_S(s,t)=\left(R_0+j\omega_s L_0\right)a_{s-t}+\frac{\delta(s-t)}{j\omega_s C_0},
	\end{equation}
	where $\delta(x)$ is the Dirac delta function. After $\mathbf{Z}_S$ is known, the admittance matrix is found by inversion: $\mathbf{Y}_S ={\mathbf{Z}^{-1}_S }$. For a given grounded substrate, we can obtain its admittance matrix $\mathbf{Y}_G$ and further the total admittance $\mathbf{Y}_{\rm tot}$, as explained in Section~\ref{Sec: Theory}. Finally, the scattering harmonics can be calculated from (\ref{Eq: Reflectance}).
	
\begin{figure}[h!]
		\centering
		\includegraphics[width=0.95\linewidth]{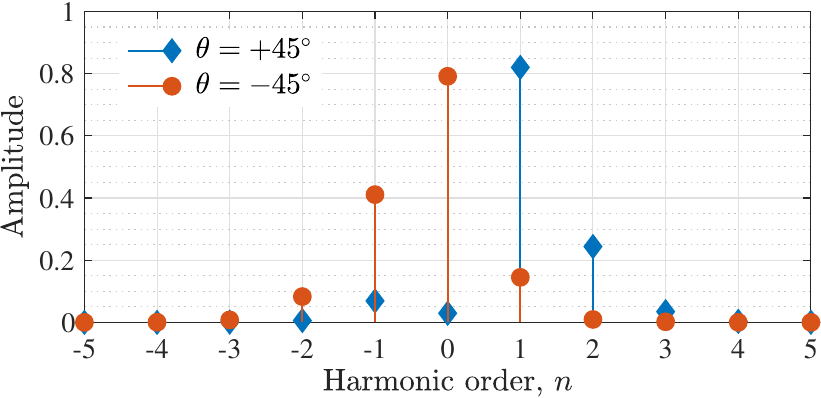}
		\caption{Distribution of harmonics at  $\omega_0=12$~THz for incidence angles $\theta=+45^\circ$ and $\theta=-45^\circ$. Here, $\beta_{\rm M}=5.86\times10^5$/m, $E_{\rm F0}=1.0$~eV, $\tau=0.5$~ps, $a_1=0.138$ and $a_2=0$. The substrate is assumed to have the parameters $d=4~\mu$m and $\epsilon_{\rm r}=4$. }\label{fig:Graphene_harmonic_comparison}
	\end{figure}

	%Lets us first study the static scenario with only space modulation ($\omega_{\rm M}=0$).
	The above theory can be used for the design of wave isolators following the ideas presented in Section~\ref{Sec: Isolators}. With a proper modulation function applied on graphene array, the incident wave can be fully converted into evanescent modes for waves from $\theta=+\theta_{\rm i}$, while for the opposite tilted angle it is mostly reflected in the specular direction. 
	To find such modulation function, we assume two unknown Fourier terms in (\ref{modulation function_graphene}), $a_1$ and $a_2$. For a fixed modulation frequency ($\omega_{\rm M}=200$~GHz) and incident angle ($\theta=+45^\circ$), a set of $a_1$,  $a_2$ and $E_{\rm F0}$ (which determines the static resistance and inductance) can be optimized to fully suppress the specular harmonic and convert the incident wave into evanescent modes.
	Figure~\ref{fig:Graphene_harmonic_comparison} shows that, at $\omega_0=12$~THz,  the wave incident from $\theta=+45^\circ$ is indeed  transformed into the first-order evanescent harmonic with a strong amplitude of $A_1=0.8$.  Swapping the incident direction invalidates the phase matching condition, and, as a consequence, the evanescent amplitudes become weak. Therefore, most of the incident energy is reflected via the specular channel.

	%For a specific medium Fermi level of graphene, one can find a proper surface impedance that excite the strong evanescent fields and totally absorb the incident wave. As shown in Fig. \ref{fig:Graphene_harmonic_comparison} , when the $E_{\rm F0}=$1.0eV, perfect absorption happens at $f=$ if $a1=$, As we decrease the Fermi level of graphene, the resonance will shift to low frequency since the inductance of  graphene $L_s$ increases with degradation of absorption performance. However, we can employ always find a proper sets of Fourier coefficients to revive the evanescent fields therefore retrieve perfect absorption. This is like a frequency tunable graphene absorber, but different with other works, here, at every frequency, the absorption can be perfect.
	
	\begin{figure}[h!]
		\centering
		\includegraphics[width=1.1\linewidth]{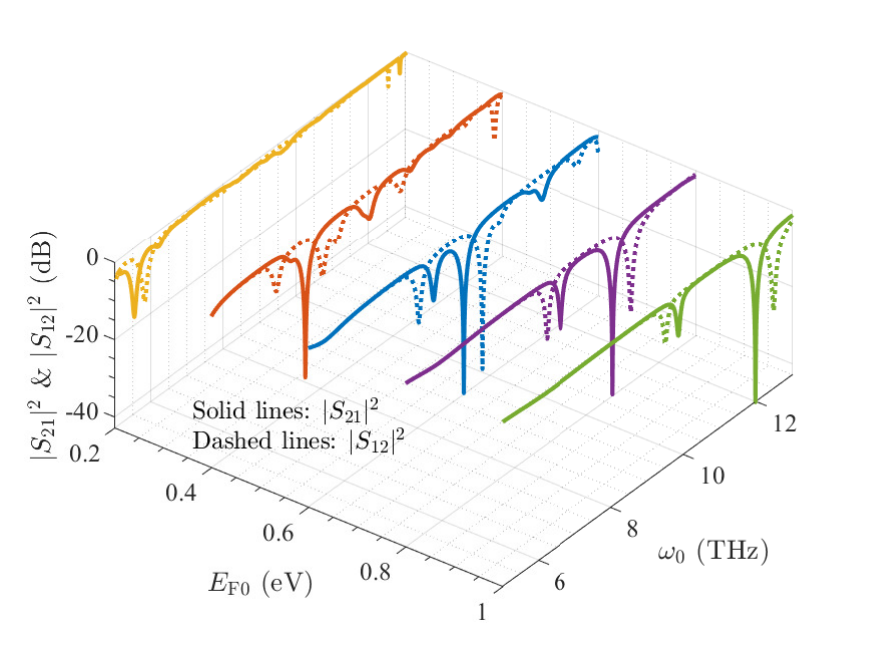}
		\caption{Frequency spectrum of the device for incidences from $\theta=+45^\circ$ (solid lines) and $\theta=-45^\circ$ (dashed lines). Green lines: $a_1=0.138$, $a_2=0$ and $E_{\rm F0}=1.0$~eV; Purple lines: $a_1=0.165$, $a_2=0$ and $E_{\rm F0}=0.8$~eV; Blue lines: $a_1=0.195$, $a_2=0$ and $E_{\rm F0}=0.6$~eV; Red lines: $a_1=0.235$, $a_2=0$,  $E_{\rm F0}=0.4$~eV; Yellow lines: $a_1=0.317$, $a_2=0.183$,  $E_{\rm F0}=0.2$~eV;}\label{fig:frequency_spectrum_graphene}
	\end{figure}
	
	Interestingly, at other frequencies, we can always find an optimal sets of $a_1$, $a_2$ and $E_{\rm F0}$ to reconstruct the isolation functionality.
	Figure~\ref{fig:frequency_spectrum_graphene} shows that the absorption frequency (from $\theta=+45^\circ$) decreases as we lower the medium Fermi level of graphene $E_{\rm F0}$. In addition, in order to achieve  good isolation performance, the modulation function $f_{\rm M}(z,t)$ should be re-optimized to maximize the absorption level. 
	Due to the modifications of the modulation function,  the isolator can  operate over a wide band from 5.5~THz to 12~THz without sacrificing the isolation performance.
	
	The isolation is determined by forward attenuation as well as by reversed insertion loss. After perfect absorption is ensured in the forward direction ($S_{21}=0$), the transmission for the opposite incidence $S_{12}$ is affected by the modulation speed as well as the resonance strength of the system which in fact determine the spectral distance between the two transmission minima. 
	The resonance of the structure can be enhanced by using low resistive (high mobility) graphene. Simultaneously, reducing the resistance of graphene also helps to relax limitations on the modulation speed since the maximum modulation frequency is determined by the intrinsic $RC$ time constant of graphene capacitors ($\omega_{\rm M}^{\rm max}=1/RC$, where $R$ and $C$ is the effective resistance and capacitance of the graphene gating) \cite{sun2016optical}. Theoretical estimations  have shown that with high quality graphene layers, the modulation speed can reach several hundred gigahertz \cite{gosciniak2013theoretical,koester2012high}.
However, in standard fabrication technology, graphene mobility significantly decreases during the fabrication process due to unavoidable impurities and contaminations from environment \cite[\textsection~1.3]{wang2014high}.  In consequence, with conventional graphene fabrication techniques, it is still a challenge to reach the modulation speed assumed in this example. On the other hand, recent advances have shown that using atomic-flat substrate (e.g., h-BN) \cite{dean2010boron}, post-annealing, and improved transfer techniques \cite{banszerus2015ultrahigh}, it is possible to obtain a high-quality graphene. 
% In addition, using metal-graphene hybrid pattern,  the effective resistance of graphene can be largely reduced \cite{wang2018toward}  to improve quality factor system.
In addition, the present electrical modulation might be replaced by an optical modulation scheme with significantly improved modulation speed \cite{li2014ultrafast,duggan2019optically}.

% For this reason, all-optical modulation schemes may be adopted for fast modulation \cite{duggan2019optically}.
% 	On the other hand,  enhancing the quality factor of the resonance also alleviates the reverse insertion loss. 
% 	This requires reduction of the effective sheet resistance of graphene, which can be achieved either by improving  graphene quality (increasing $\tau$) in the fabrication process or employing metal-graphene hybrid structures \cite{wang2018toward}.

	\section{Conclusions}
	
In this paper, we have introduced a general approach to realize nonreciprocal metasurface devices to control waves in single or multiple scattering channels in free space. Importantly, we have shown that a variety functionalities can be realized using the same physical platform by adjusting the space-time modulation of the single tunable element: an impedance sheet.
As examples, the proposed technique has been applied to design  nonreciprocal metasurfaces which act as isolators, nonreciprocal phase shifters, quasi-isolators, and circulators. 
In addition, a graphene-based modulation platform has been proposed, where the isolation frequency can be dynamically tuned in a wide frequency range by changing the modulation function.
We expect that the introduced technique can be useful in the developments of future computer-controlled intelligent  nonreciprocal metasufaces for various applications.

	\section{Acknowledgments}
	
	This work received funding from the Foundation for Aalto University Science and Technology, the Academy of Finland (project 309421), the European
	Union’s Horizon 2020 research and innovation programme -- Future Emerging Topics (FETOPEN) under grant agreement No 736876, and the U.S. Air Force Office of Scientific Research. The authors would thank Dr.~Viktar Asadchy, Dr.~Dimitrios Sounas, Dr.~Fu Liu, and Dr.~Beibei Kong for helpful discussions.

	\bibliography{references}

\end{document}